\documentclass[12pt]{article}
\usepackage[latin9]{inputenc}
\usepackage{amsmath}
\usepackage{amssymb,color}
\usepackage{graphicx}
\usepackage{esint}
 \usepackage{cite}
 
\makeatletter
\usepackage{subfigure}\usepackage{amscd}
\usepackage{appendix}
\usepackage{verbatim}
\usepackage{epstopdf}

\usepackage{color}

\title{Monopolium}

\begin{document}
	\begin{titlepage}
		\hspace{0.75cm}{\hbox to\hsize{\hfill{} CTPU-PTC-21-13, OCHA-PP-365}}
		
		\bigskip{}
		\vspace{3\baselineskip}
		
		\begin{center}
			\textbf{\Large{Searching for Monopoles via Monopolium Multiphoton Decays}}\par
		\end{center}{\Large \par}
		
		\vspace{0.1cm}
		\begin{center}
			\textbf{
				Neil D. Barrie$^{1\dagger}$, Akio Sugamoto$^{2\ddagger}$, Matthew Talia$^{3\S}$, and Kimiko Yamashita$^{4\P}$
			}\\
			\textbf{ }
			\par\end{center}
		
		\begin{center}
			{\it
				$^1$ Center for Theoretical Physics of the Universe, Institute for Basic Science (IBS), Daejeon, 34126, Korea\\
				${}^{2}$Department of Physics, Graduate School of Humanities and Sciences,\\
				Ochanomizu University, 2-1-1 Ohtsuka, Bunkyo-ku, Tokyo 112-8610, Japan\\
				${}^{3}$Astrocent, Nicolaus Copernicus Astronomical Center Polish Academy of Sciences, ul. Bartycka 18, 00-716 Warsaw, Poland\\
				${}^{4}$ Institute of High Energy Physics, Chinese Academy of Sciences, Beijing 100049, China \\
				${}^{\dagger}$nlbarrie@ibs.re.kr,
				${}^{\ddagger}$sugamoto.akio@ocha.ac.jp,
				${}^{\S}$ mtalia@camk.edu.pl,
				${}^{\P}$ kimiko@ihep.ac.cn}
			\textit{\small{}}
			\par\end{center}{\small \par}
		
		\noindent 
		\begin{center}
			\textbf{\large{}Abstract}

			\par\end{center}{\large \par}

		We explore the phenomenology of a model of monopolium based on an electromagnetic dual formulation of Zwanziger and lattice gauge theory. The monopole is assumed to have a finite-sized inner structure based on a 't Hooft-Polyakov like solution, with the magnetic charge uniformly distributed on the surface of a sphere. The monopole and anti-monopole potential becomes linear plus Coulomb outside the sphere, analogous to the Cornell potential utilised in the study of quarkonium states. Discovery of a  resonance feature in the diphoton channel as well as in a higher multiplicity photon channel would be a smoking gun for the existence of monopoles within this monopolium construction, with the mass and bound state properties extractable. Utilising the current LHC results in the diphoton channel, constraints on the monopole mass are determined for a wide range of model parameters. These are compared to the most recent MoEDAL results and found to be significantly more stringent in certain parameter regions,  providing strong motivation for exploring higher multiplicity photon final state searches.
		
		\noindent 
		
	\end{titlepage}
	
	\noindent 
	\section{Introduction}
	
	The existence of magnetic monopoles was first postulated by Dirac, with their existence found to provide an explanation for the quantisation of electric charge \cite{Dirac1,Dirac2,Dirac3}. Monopoles have a rich phenomenology with connections to new physics scales as they are generic predictions of spontaneously broken Grand Unified Theories (GUT). Thus, discovering monopoles would provide a window through which to explore higher energy scales and new physics both terrestrially and cosmologically. Possible early universe phase transitions associated with them could also lead to gravitational waves, providing an additional avenue for discovery. There are  important cosmological implications for monopoles due to their stability, ensured by magnetic charge conservation, such as electroweak Baryogenesis and Dark matter \cite{Preskill:1979zi,Arunasalam:2017eyu,Arunasalam:2018iom,Vento:2020vsq}.

	Monopoles have been extensively searched for experimentally, but have yet to be discovered \cite{Patrizii:2015uea,Mavromatos:2020gwk}. One important way to search for monopoles is through production at colliders which are able to probe monopoles of low mass relative to that usually expected from GUT models, such as current efforts by CMS, ATLAS, and MoEDAL collaborations. No evidence of monopoles has been found to date, with the current collider lower bounds on the monopole mass reaching as high as a few TeV, with dependence upon the nature of the monopoles \cite{Acharya:2019vtb,Aad:2019pfm}. These direct production searches face many technical and theoretical difficulties due to monopoles strong couplings and their highly ionising nature.

	One exciting prospect for observing monopoles is through monopole-anti-monopole bound states, known as monopolium, and their associated decay products 
	\cite{Nambu,sugamoto,Epele1, Epele:2008un, Epele2,Yamadaetal,Barrie:2016wxf,Epele:2016wps,Saurabh:2017ryg,Fanchiotti:2017nkk,Reis:2017rvb,Fredericks:2017mmg,Vento:2018sog,Baines:2018ltl,Heras:2019nhv,Vento:2019auh,Abreu:2020qqy}. The monopolium decay to diphotons provides a clean signal region in which to observe possible decays, compared to the types of signals expected from monopole-anti-monopole pair production. Additionally, the potential large negative binding energy of the monopolium states opens the window to probing higher energy scales beyond that which is possible through direct production at colliders. The large binding energy means the production of monopolium is highly favoured over pair production, while allowing a probe of the higher energy physics scale of the constituent monopoles. Such searches can provide complementary avenues for discovery to those done at MoEDAL and other collider searches, and illuminate the dynamics of the monopoles and their possible bound state formation. The details of the monopolium bound state  are not clearly understood because of the strong couplings and subsequent non-perturbative nature. Various analyses has been done, but many uncertainties remain with mainly qualitative results able to be gleaned \cite{Epele1,Epele:2008un,Epele2,Barrie:2016wxf}. By considering cross-section constraints in the diphoton channel, it may be possible to derive constraints on the monopolium states \cite{Dougall:2007tt,Sirunyan:2018wnk,Aad:2021yhv}. In future, the measurement of higher multiplicity photon final states may provide additional avenues for discovery because the  Standard Model background is expected to be small, and the strong coupling of the monopoles to photons potentially leading to large signals.

	In a previous work, we presented a possible method for constructing a model of a monopole-anti-monopole bound state \cite{Barrie:2016wxf}. The model treated the monopolium analogously to a quarkonium state, deriving a modified Cornell potential \cite{Eichten:1978tg}, while taking into account the possible shielding of the internal structure of the monopole at a finite scale. We utilise a strong coupling expansion in lattice gauge theory, and accordingly a linear term is included in the monopole-anti-monopole potential, which allows for large binding energies to be studied. The phenomenological aspect of our work  was initially motivated by the tentative evidence for a new resonance in the diphoton channel and the possible bound state nature \cite{Neil1,Neil2,Barrie:2017eyd}, and thus had a narrow phenomenological scope with  the main focus the construction of the bound state.  In this work, we make improvements on various aspects in the construction and investigation of the model, and provide a more thorough exploration of the possible collider phenomenology within this framework.

	The paper is structured as follows; firstly, in Section \ref{monopolium} we briefly review the monopole-anti-monopole bound state model proposed in our previous work \cite{Barrie:2016wxf}. In Section \ref{boundstate}, we will discuss the properties of the monopolium state, including the binding energy, bound state wave function, and the diphoton decay properties. Section \ref{twophoton_only} derives the expected diphoton cross section for the monopolium states at the LHC, including the possible contributions of higher energy eigenstates. The implications to the collider phenomenology upon including higher multiplicity photon decays is  explored in Section \ref{multiphoton}. After which we present a discussion of the results and concluding remarks, including comments on future prospects.


	\section{Model of Monopolium Bound State}
	\label{monopolium}
	
	In our previous work, we introduced a model of monopolium based on the Zwanziger formalism. The resultant potential for the monopolium state, which we shall investigate in this work, is analogous to the Cornell potential utilised in studies of quarkonium states \cite{Eichten:1978tg}.
	We assume that there is only one kind of spin 1/2 monopole $M$ in our world, that is electrically neutral, having magnetic charge $g$.  Within the Standard Model we have a number of magnetically neutral spin 1/2 fields with electric charge $Q=\pm e$  ($e$ is the unit of the electron charge), and the monopole satisfies the Schwinger quantization condition with these charged particles.  This gives
	\begin{eqnarray}
	g=\frac{4\pi}{e} N ~~(N=\mbox{integer}) ~.  
	\end{eqnarray}
	
	The difficulty of highly charged particles is the estimation of the potential between them.  One photon exchange is insufficient, and so we will use a lattice gauge theory approach with a finite lattice constant $a$, in which a strong coupling expansion is possible \cite{lattice_gauge_theory1, lattice_gauge_theory2, lattice_gauge_theory3}.  We will apply this to the magnetic $ U(1) $ part of the manifestly electromagnetic dual formulation of Zwanziger \cite{Zwanziger1, Zwanziger2}. However,  $ U(1) $ gauge theories are not well defined in lattice gauge theory, due to the existence of a first-order phase transition between the weak coupling perturbative region and the strong coupling confinement region, and so the continuum limit of $a \to 0$ cannot be consistently taken \cite{Creutz-Jacobs-Rebbi}.
	On the other hand, the continuum limit is properly taken for $ SU(2) $ \cite{Creutz} and other asymptotically free gauge theories.  So, we consider that a non-Abelian structure is revealed when we go inside the finite sized $ U(1) $ monopole, with the gauge group expected to be enhanced from $ U(1) $ to $ SU(2) $ as $a \to 0$, and the 't Hooft-Polyakov-like structure will appear \cite{monopole1, monopole2}.  In the 't Hooft-Polyakov monopole,  the $ U(1) $ magnetic charge is located at the origin as a point-like singularity.  Instead, we consider that the $ U(1) $ magnetic charge is distributed non-locally, with magnetic charge  distributed uniformly on the surface of a sphere with radius $R$, for which a solution exists.  Inside the sphere ($r<R$) there is no magnetic force, and so the potential becomes flat, while the potential between the monopole and anti-monopole becomes linear plus Coulomb outside the sphere.   Below we provide a summary of the details of the monopolium construction postulated in our previous work.

	\subsection{Zwanziger's manifestly dual formulation of gauge theory} 
	The manifestly dual formulation of $ U(1) $ gauge theory by Zwanziger \cite{Zwanziger1, Zwanziger2} is  given by the following action,   
	\begin{eqnarray}
	S^{\rm{ZW}}&=&\int d^4 x \left[ \left(\frac{1}{2} \eta^{\mu} \eta^{\lambda}
	\sqrt{-g} g^{\nu\rho}\right) \nonumber \right. \\
	&\times& \left(F_{\mu\nu} F_{\lambda\rho} + G_{\mu\nu} G_{\lambda\rho}+ F_{\mu\nu} \tilde{G}_{\lambda\rho} - G_{\mu\nu}\tilde{F}_{\lambda\rho} \right)  \nonumber \\
	&+& \left. \sum_i \overline{\psi_i} \gamma^{\mu}( iD_{\mu}-m_i) \psi_i \right]~.
	\end{eqnarray}
	Here, a constant unit vector $\eta^{\mu}$, denoting the direction of Dirac strings \cite{Dirac1, Dirac2, Dirac3}, is displayed in parallel along the space-like direction, and
	\begin{eqnarray}
	iD_{\mu}&=&i \partial_{\mu}- e_i A_{\mu} -g_i B_{\mu}~, \\
	F_{\mu\nu}&=&\partial_{\mu}A_{\nu}-\partial_{\nu}A_{\mu}~, ~~
	G_{\mu\nu}=\partial_{\mu}B_{\nu}-\partial_{\nu}B_{\mu}~, \\
	\tilde{F}_{\mu\nu}&=&\frac{1}{2} \varepsilon_{\mu\nu\lambda\rho}
	F^{\lambda\rho}, ~~\tilde{G}_{\mu\nu}=\frac{1}{2} \varepsilon_{\mu\nu\lambda\rho}
	G^{\lambda\rho}~, 
	\end{eqnarray}
	where fermions with electric charge $e_i$ and magnetic charge $g_i$ are introduced, and $\varepsilon^{0123}=1$. From the consistency condition of the finite Lorentz transformation, the following Dirac, or Schwinger type, quantization condition \cite{Dirac1, Dirac2, Dirac3} appears,
	\begin{eqnarray}
	e_ig_j-g_ie_j= 4\pi N_{ij}~,
	\end{eqnarray}
	where $N_{ij}$ is an integer.
	Here, we consider a flat space-time $g_{\mu\nu}=\eta_{\mu\nu}=(1, -1, -1, -1)$ and $\varepsilon^{0123}=-\varepsilon_{0123}=1$.    In the Zwanziger formulation, the degrees of freedom are doubled by the introduction of electric and magnetic vector potentials, but are halved by the projection to the $\eta^{\mu}$ direction.
	
	If the axial gauge is taken,
	\begin{eqnarray}
	\eta^{\mu}A_{\mu}(x)=\eta^{\mu}B_{\mu}(x)=0~,
	\end{eqnarray}
	no ghost fields appear, and the Feynman rules are obtained as follows,
	\begin{eqnarray}
	\langle A_{\mu} A_{\nu} \rangle (k)&=& \frac{-i}{k^2 + i \varepsilon} \left( g_{\mu\nu} - \frac{k_{\mu}\eta_{\nu}+k_{\nu}\eta_{\mu}}{k \cdot \eta}+\eta^2 \frac{k_{\mu}k_{\nu}}{(k \cdot \eta)^2} \right)~, \\
	\langle B_{\mu} B_{\nu} \rangle (k)&=& \frac{-i}{k^2 + i \varepsilon} \left( g_{\mu\nu} - \frac{k_{\mu}\eta_{\nu}+k_{\nu}\eta_{\mu}}{k \cdot \eta} +\eta^2 \frac{k_{\mu}k_{\nu}}{(k \cdot \eta)^2}\right)~, \\
	\langle A_{\mu} B_{\nu} \rangle (k) &=&-\langle B_{\mu} A_{\nu} \rangle  (k)= \frac{-i}{k^2 + i \varepsilon} \varepsilon_{\mu\nu\rho\sigma} \frac{\eta^{\rho}k^{\sigma}}{k \cdot \eta}~.
	\end{eqnarray}
	The kinetic terms of the gauge field are complicated and depend on $\eta^{\mu}$, but  the $2 \times 2$ matrix form of the propagators is simple and satisfies,
	\begin{eqnarray}
	\hat{D}^{ab\mu\nu} \langle V_{b\nu} V_{c\lambda} \rangle = \delta^a_c \eta^{\mu}_{\lambda}~,
	\end{eqnarray} 
	where $V^1_{\mu}=A_{\mu}$, $V^2_{\mu}=B_{\mu}$, and $\hat{D}^{ab\mu\nu}$ is the differential operator for the gauge fields in the action.

	\subsection{Monopolium Formulation}

	The bound state of the monopoles is formed by the exchange of magnetic photons $\gamma_M$.  However, the coupling of the monopole to the magnetic photon is very strong, so that we adopt a lattice gauge theory approach with a lattice constant $a$ as a UV cutoff \cite{lattice_gauge_theory1, lattice_gauge_theory2, lattice_gauge_theory3}.
	The space-time is considered to be a square lattice $n=(n_0, n_1, n_2, n_3)$,  where $n_{\mu}$ are integers, with a lattice constant $a$. The link variables $U^{(A)}_{n\hat{\mu}}$ and $U^{(B)}_{n\hat{\mu}}$ are introduced as usual for the electric and magnetic photons $A_{\mu}(x)$ and $B_{\mu}(x)$, respectively,
	\begin{eqnarray}
	U^{(A)}_{n\hat{\mu}}=e^{i \{e a A_{\mu}(na)\}}, ~~U^{(B)}_{n\hat{\mu}}=e^{i \{g a B_{\mu}(na)\}}~,
	\end{eqnarray}
	and the Wilson loops $W^{(A)}[C]$ and $W^{(B)}[C]$ are defined as the product of the link variables along the loop $C$:
	
	\begin{eqnarray}
	W^{(A)}[C]= \prod_{n \in C, ~\mu \parallel C} U^{(A)}_{n\hat{\mu}}, ~~W^{(B)}[C]= \prod_{n \in C, ~\mu \parallel C} U^{(B)}_{n\hat{\mu}}~.
	\end{eqnarray}
	The minimum Wilson loop is given for the boundary curve $C_{n\mu\nu}\equiv \partial P_{n\mu\nu}$ of the minimum region $P_{n\mu\nu}=\mbox{rectangular}~(n, n+\hat{\mu}, n+\hat{\mu}+\hat{\nu}, n+\hat{\nu}, n)$. Then, the Zwanziger action can be written as the lattice gauge theory action in the Euclidean metric,
	\begin{eqnarray}
	S^{\rm{ZW}}_\mathrm{lattice}&=&-\sum_{n, \nu} \left( \frac{1}{2e^2} W^{(A)} [C_{n\eta\nu}] +\frac{1}{2g^2} W^{(B)} [C_{n\eta\nu}]+ (h.c.) \right) \nonumber \\
	&-&\sum_{n, \nu} \frac{1}{2eg}  \left(W^{(A)} [C_{n\eta\nu}]\tilde{W}^{(B)} [C_{n\eta\nu}] - W^{(B)} [C_{n\eta\nu}]\tilde{W}^{(A)} [C_{n\eta\nu}] \right) \nonumber \\
	&+&\frac{a^3}{2} \sum_{i, n\nu} \left( \overline{\psi}_{i, n}\gamma_{\nu} \left(U^{(A)}_{n\nu}+U^{(B)}_{n\nu}\right) \psi_{i, n+\nu}-\overline{\psi}_{i, n}\gamma_{\nu} \left(U^{(A)}_{n\nu}+U^{(B)}_{n\nu}\right)^{\dagger} \psi_{i, n-\nu} \right) \nonumber \\
	&-& a^4 \sum_{i, n} m_i ~\overline{\psi}_{i, n} \psi_{i, n} ~,
	\end{eqnarray}
	where the dual Wilson loop reads
	\begin{eqnarray}
	\tilde{W}^{(A,B)} [C_{n\eta\nu}] =-\frac{i}{2} \epsilon_{\eta\nu\lambda\rho}W^{(A,B)} [C_{n\lambda\rho}] ~.
	\end{eqnarray}
	
	In the action, we assume that the magnetic coupling $g$ is strong, while the electric coupling $e$ is weak and perturbative.  So, in estimating the expectation value of the large Wilson loop $W^{(B)}[C]$, the strong coupling expansion is used \cite{lattice_gauge_theory1, lattice_gauge_theory2, lattice_gauge_theory3}.  We choose $C$ to be a rectangle of length $T$ in time and length $r$ in the space-like direction $\mu$; then we have 
	\begin{eqnarray}
	\langle W^{(B)}[C] \rangle &=& \frac{\int dU^{(B)}_{n\nu}~ W^{(B)}[C] ~e^{-S^{\rm{ZW}}_\mathrm{lattice}}}{\int dU^{(B)}_{n\nu} ~e^{-S^{\rm{ZW}}_\mathrm{lattice} } }\\
	&=& \delta_{\mu\eta} \exp\left(-\ln(2g^2) \frac{T \cdot r}{a^2} \right) \left(1+ \cdots \right)~,
	\end{eqnarray}
	where $T \cdot r$ denotes the minimum area of the rectangle $C$, so that the Wilson's area law is realized only if $\mu$ is in the $\eta$-direction. 
	
	We define the potential between a heavy monopole and its anti-monopole separated by a distance $r$ to be $V(r)$.  Then, the potential has a linear term in $r$, if the monopole and anti-monopole are separated in the $\eta$-direction;
	\begin{eqnarray}
	V(r)= \delta_{r \parallel \eta} \frac{\ln(2g^2)}{a^2} r + \cdots~.
	\end{eqnarray}
	
	This clarifies the meaning of the special direction $\eta$ that appears in the Zwanziger formulation, giving the direction of the Dirac string starting from the monopole.  Therefore, the monopole and anti-monopole are connected by the string, starting from the monopole and ending at the anti-monopole, which contributes to the linear potential between them.
	
	In addition to this strong coupling contribution, we will add the usual perturbative weak coupling contribution, and so the potential at this stage is
	\begin{eqnarray}
	V(r)= -\frac{g^2}{4\pi r} + \frac{\ln(2g^2)}{a^2} r~.
	\end{eqnarray}
	
	This matches with high precision QCD calculations in which the potential between a quark and anti-quark pair is well approximated by the linear plus Coulomb potential \cite{Bali1, Bali2, Bali3}; this potential is known as the Cornell potential and is well-defined for point-like quarks.  The monopole, however, may not be point-like, and may have an internal structure.  The $ U(1) $ lattice gauge theory is usually considered not to be well defined, since there exists a first-order phase transition between the confinement phase and the perturbative phase \cite{Creutz-Jacobs-Rebbi}, and it obstructs the continuum limit of $a \to 0$.  One way out from this difficulty is to lift the $ U(1) $ theory to $ SU(2) $ theory, or other asymptotically free theory, when we approach the short distance region.  As is shown by 't Hooft and Polyakov \cite{monopole1, monopole2}, the $ U(1) $ monopole was given as a classical solution of $ SU(2) $ gauge theory, which is broken to a $ U(1) $ by a triplet Higgs field $\phi^a(x)~(a=1,~2,~3)$. Therefore, if we go inside the monopole, the non-Abelian gauge theory may appear, in which case, we may take the continuum limit properly.  
	
	The 't Hooft-Polyakov monopole is a classical solution of the $ SU(2) $ gauge theory with a triplet Higgs, based on,
	\begin{eqnarray}
	{\cal L}_{SU(2)\, \mathrm{monopole}}=-\frac{1}{4} (F^a_{\mu\nu})^2+ (D_{\mu}\phi^a)^2-\lambda(\vert\phi \vert^2-v)^2~,
	\end{eqnarray}
	and the following ansatz:
	\begin{eqnarray}
	A^a_i(x)=v~\epsilon^{aij}\hat{r}^j \frac{1-K(\xi)}{\xi}, ~\phi^a(x)=v~\hat{r}^a \frac{H(\xi)}{\xi}~,
	\end{eqnarray}
	where we define a dimensionless parameter $\xi=evr$.
	If we take the limit $\lambda \to 0$ while keeping $v \ne 0$, the solution, called the BPS solution, is given analytically \cite{BPS1, BPS2}.  Then, the equations of motion become  first order,
	\begin{eqnarray}
	\xi \frac{dK}{d\xi}=-KH, ~~\xi \frac{dH}{d\xi}=H-K^2+1~,
	\end{eqnarray}
	which have the following solutions,
	\begin{eqnarray}
	K(\xi)=\frac{\xi}{\sinh \xi}, ~~H(\xi)=\xi \coth\xi-1~.
	\end{eqnarray}
	
	We want to know the distribution of the magnetic charge inside the monopole, $\xi <1$ or $r< 1/ev$.  The $ SU(2) $ gauge potential and the Higgs field are properly reduced by factors of $1-K(\xi)$ and $H(\xi)$, but the monopole charge is unfortunately not smeared even inside the monopole.  This can be understood from the $ U(1) $ field strength proposed by 't Hooft.  This gauge invariant expression can be rewritten as follows \cite{Arafune-Freund-Goebel},
	\begin{eqnarray}
	F_{\mu\nu}=\partial_{\mu} (\hat{\phi}^a A^a_{\nu})-\partial_{\nu} (\hat{\phi}^a A^a_{\mu})-\frac{1}{e}\epsilon^{abc}\hat{\phi}^a \partial_{\mu} \hat{\phi}^b \partial_{\nu} \hat{\phi}^c~,
	\end{eqnarray}
	where $\hat{\phi}^a=\phi^a/\vert \phi \vert$.  From this expression the magnetic charge is found to be a topological number,
	\begin{eqnarray}
	\Phi_m=\int \textbf{B}d\textbf{S}=\frac{4\pi}{e} \int d^3x \frac{\partial(\phi^1, \phi^2, \phi^3)}{\partial(x^1, x^2, x^3)}=\frac{4\pi}{e} N~,
	\end{eqnarray}
	where $\Phi_m$ is the magnetic flux, and $N$ is the winding (wrapping) number of the sphere of the Higgs fields ($\boldsymbol\phi^2=v^2$)  by the  sphere in space ($\textbf{x}^2=1$).
	
	We will consider the scenario in which the magnetic charge is distributed uniformly on the surface of the sphere,  that is one of the solutions that exists.  Let's take the following modified gauge and Higgs fields in which the contribution from $r<R$ is cutoff,
	\begin{eqnarray}
	\tilde{A}^a_i(x)=\theta(r-R) A^a_i(x), ~\tilde{\phi}^a(x)=\theta(r-R)\phi^a(x)~,
	\end{eqnarray}
	where $\theta(r-R)$ is a step function $0$ for $r<R$, and $1$ for $r>R$.  Accordingly, $K$ and $H$ are modified:
	\begin{eqnarray}
	1-\tilde{K}(r)=\theta(r-R) (1-K(r)), ~~\tilde{H}(r)=\theta(r-R)H(r)~.
	\end{eqnarray}
	
	This modification inside the sphere ($r<R$) is allowed, since $K(r)=1$ and $H(r)=0$ is the solution.
	
	The distribution of the $ U(1) $ magnetic charge is easily understood. Since $\phi^a=0$ inside $r<R$, the magnetic flux on the surface of the radius $r$ sphere reads,
	\begin{eqnarray}
	\Phi_m(r)= \frac{4\pi}{e} \theta(r-R)~,
	\end{eqnarray}
	which shows that the magnetic charge $4\pi/e$ is distributed uniformly on the surface of the monopole sphere with radius $R$.  The solution satisfies the following equations,
	\begin{eqnarray}
	\xi \frac{d\tilde{K}}{d\xi}&=&-\tilde{K}\tilde{H} + evR\left(1-K(evR)\right)\delta(\xi-evR)~, \\
	\xi \frac{d\tilde{H}}{d\xi}&=&\tilde{H}-\tilde{K}^2+1+evRH(evR)\delta(\xi-evR)~,
	\end{eqnarray}
	which are modified only on the surface, where the magnetic charge is distributed.
	
	Therefore, the magnetic force between the monopole and anti-monopole vanishes when one goes inside either respective sphere $(r<R)$. We will consider that the linear potential is the dominant contribution and the Coulomb potential plays the role of lowering the potential. Then, the corresponding potential is,
	\begin{equation}
	V(r)= \begin{cases}
	V_1(r)=\mbox{const} = - \frac{\alpha_g}{ R} + \frac{\ln(8 \pi \alpha_g)}{a^2} R~,              & (\mbox{for}~ r<R)~,\\
	V_2(r)=- \frac{\alpha_g}{ R} + \frac{\ln(8 \pi \alpha_g)}{a^2} r~,& (\mbox{for}~ r>R)~,
	\end{cases}
	\end{equation}
where $ \alpha_g=\frac{g^2}{4 \pi}=\frac{1}{\alpha}=137 $ .	We shall utilise this potential to describe the monopolium bound state in what follows.


	\section{Wave Function and Binding Energy  of Monopole in Bound State} 
	\label{boundstate}
	
	Using the potential $V(r)$, we can obtain the wave function and energy eigenvalues of the monopolium states.  Given that we have a piecewise potential, we solve the Schr\"{o}dinger's equation requiring continuity at the point $ r=R $ in the radial wavefunction, where we will assume spherical symmetry. We define the potential as follows and derive the radial wavefunction for both potentials,
	\begin{eqnarray}
	V_1(r) &=& - \frac{137}{ R} + \kappa  R ~~ (\mbox{for}~ r<R)~, \label{pots1} \\
	V_2(r)&=& - \frac{137}{ R} + \kappa  r~~~(\mbox{for}~ r>R)~,
	\label{pots2}
	\end{eqnarray}
	where we denote the string tension by $\kappa=\ln(8 \pi \alpha_g)/a^2$.
	
	The radial wavefunction is defined by  $\psi(r)= \chi(r)/r$, such that $\chi(r)$ satisfies the one-dimensional Schr\"{o}dinger equation. From the potentials given in Eqs. (\ref{pots1}) and (\ref{pots2}), the corresponding radial wavefunction solutions are described by sinusoidal and Airy functions, respectively,
	\begin{eqnarray}
	\chi_1(r)&=& C_1\sin\left(m r\sqrt{2x_n/m}\right)~~~~~~~~~~~~~~~~~ (\mbox{for}~ r<R)~, \\
	\chi_2(r)&=& C_2 \textrm{Ai}\left[2\left( \frac{m^2}{2\kappa } \right)^{\frac{2}{3}}\frac{\kappa (r-R)-x_n}{m}\right]~~(\mbox{for}~ r>R)~,
	\end{eqnarray}
	where the boundary conditions $ \chi_1(0)=0 $ and $ \lim\limits_{r\rightarrow \infty}\chi_2(r)=0 $ have been imposed, and the $ n $th energy eigenvalue is,
	\begin{eqnarray}
	\Delta E_n= - \frac{137}{ R} + \kappa  R +x_n~,
	\end{eqnarray}
	with corresponding eigenstate mass,
	\begin{equation}
	M_n = 2 m - \frac{137}{R}+ \kappa R  +x_n~,
	\end{equation}
	for $ n\geq 1 $~, where $ x_n=x_n(p,q,m) $ is a function of $ p $, $ q $ and $ m $, that is to be determined numerically.
	
	In order to determine $ C_1 $ and $ C_2 $, and the corresponding energy eigenvalues, we require that the following three conditions are satisfied,
	\begin{eqnarray}
	\chi_1(R)=\chi_2(R) ~~\textrm{and} ~~ \frac{d}{dr}\chi_1(R)=\frac{d}{dr}\chi_2(R) ~,
	\end{eqnarray}
	for continuity, and 
	\begin{eqnarray}
	4 \pi \int_0^{\infty}  \vert \chi (r) \vert^2 dr =4 \pi \left(\int_0^{R}  \vert \chi_1 (r) \vert^2 dr +\int_R^{\infty}  \vert \chi_2 (r) \vert^2 dr\right) = 1~,
	\end{eqnarray}
	to normalise the wavefunction. These can be solved numerically, and the energy eigenvalues derived. By requiring $ 0<M_n<2m $, the list of possible spherically symmetric bound state can be compiled for given input parameters, $ R$, $a $ and $ m $. 
	
	In the analysis that follows we will make the identification that $ R=\frac{q}{m} $ and $ a=\frac{p}{m} $. The energy scales in this scenario are given by both the core size of the monopole, and necessarily when the QED description would break down embodied in the lattice scale $ a $. Furthermore, motivated by the usual classical monopole radius we will assume that $ R $ and $ a $ are proportional to $ \frac{1}{m} $, with  constants of proportionality defined as $ q $ and $ p $  \cite{Preskill:1984gd}. The most motivated choice of $ q $ is 137 from the definition of the approximate monopole core radius, there may be some ambiguity in this denominator so we will allow it to vary for our analysis. In this parametrisation, the monopole mass parameter $ m $ factors out for many of the calculations, allowing determination of the monopolium properties by varying the parameters $ q $ and $ p $, each exhibiting unique bound state signatures and phenomenology.
	
	We consider that the monopole and anti-monopole collide when the distance $r=R$ in our model, so that the wave function is evaluated at $r=R$.

\begin{figure}[h!]
		\centering
		\includegraphics[width=0.5\textwidth]{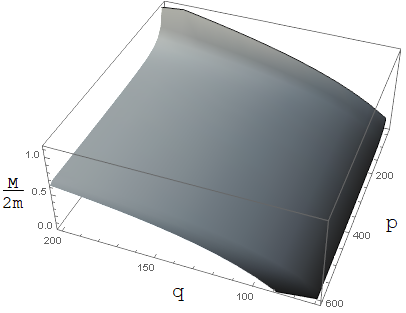}
		\caption{Lower bound of the monopolium to monopole mass ratio for all mass eigenvalues, for given $ q $ and $ p $. }
		\label{Mx}
	\end{figure}

	The eigenvalue masses are determined by the value of $ x_n $, which increases with $ n $. That is,
	\begin{equation}
	M_n = 2 m - \frac{137}{R}+ \kappa R  +x_n\simeq \left(2-\frac{137}{q}+\frac{q\ln(1096\pi) }{p^2}\right)m +x_n ~,
	\label{mon_mass}
	\end{equation}
	for $ n\geq 1 $, and subsequently the lower bound on the monopolium mass $ M_* $, for a given $ q $, is given by,
	\begin{equation}
	M_0 = 2 m - \frac{137}{R}+ \kappa R\simeq \left(2-\frac{137}{q}+\frac{q\ln(1096\pi) }{p^2}\right)m ~,
	\end{equation}
	and is depicted as a function of $ q $ and $ p $ in Figure \ref{Mx}. For large $ p $, the smallest allowed $ q $ value can be derived to be approximately $ 68.5<q $, above which $ M_n>0 $ is always satisfied. Although smaller $ q $ values can be accessed by choosing $ q>p $, in what follows we explore $ q\geq 70 $ to ensure that the monopolium masses considered always satisfy $ M_0>0 $. We will require $ 2m>M_n>0 $, for which the monopolium states exhibit negative binding energies, opening the possibility to probing energy scales much greater than production of the constituent particles would allow, particularly for large binding energy scenarios.
	
	In the following we will confine our analysis to $ q\in[70,200] $ and $ p\in[60,2000]  $, which takes into all major regions of interesting behaviour in this model. $ p $ values that are too small can lead to monopolium masses greater than $ 2m $, while for $ q $ this would give $ M<0 $. We do not consider combinations in which $ q $ is so small and $ p $ so large that $ M\rightarrow 0 $, which we ensure be considering $ q>70 $. The large negative binding energies we obtain  allow us to probe monopole masses well beyond those probed by pair production.

	Now we can investigate the properties of the wavefunction for the corresponding energy eigenvalues. Upon numerically solving the above equations, the number of eigenvalues obtained in the region $ M_0<M<2m $, for the range of $ q $ and $ p $ parameters to be considered, varies between 100-600 states. Below we will also consider the stability and decay properties of these different energy level states, and the phenomenological implications.


	\subsection{ Diphoton decay width of Monopolium States}

	After obtaining the energy and wave function of the monopolium, we can estimate the decay properties of the monopolium states. The decay rate of of the monopolium state to diphotons  can be estimated from the cross section $\sigma(m+\overline{m} \to 2\gamma)$ as,
	\begin{equation}
	\Gamma(M \to 2\gamma)=4 \rho \sigma(m+\overline{m} \to \gamma + \gamma) v_{rel}~,
	\end{equation}
	where
	\begin{equation}
	\rho =\frac{\int_{0}^{R}\vert \psi(r)\vert^2 dr}{\frac{4}{3}\pi R^3}~,
	\end{equation}
	since the monopoles have a finite core size $R$, we use the average probability density of finding the monopole and anti-monopole at a relative distance $ R $ or less in the bound state. We assume that a distance of $ r\leq R $ indicates collision of the monopole-anti-monopole pair, and hence annihilation.  The factor 4 $\sigma v_\mathrm{rel}$ gives the reaction rate, with the factor 4 coming from the possible combination of monopole and anti-monopole spin states. 
	
	In this calculation, the polarization vector of the magnetic photon $\epsilon_m(k)^{\mu}$, has to be converted to the usual electric photon's polarization $\epsilon_{\gamma}(k)^{\nu}$, by using the off-diagonal propagator $\langle B_{\mu} A_{\mu} \rangle (k)$, that is,
	\begin{eqnarray}
	\epsilon_m(k)^{\mu}=\varepsilon^{\mu\nu\rho\sigma} \epsilon_{\gamma}(k)_{\nu} \frac{\eta_{\rho} k_{\sigma}}{(k \cdot \eta)}~, \label{replacement of polarizations}
	\end{eqnarray}
	The polarization sum of photons, $\sum_{pol}\epsilon_{\gamma}(k)^{*\lambda}\epsilon_{\gamma}(k)^{\rho}=-g^{\lambda\rho}$, is performed, then we have
	\begin{eqnarray}
	\sum_{\gamma's~\mathrm{pol.}}\epsilon_m(k)^{*\mu}\epsilon_m(k)^{\nu}
	=g^{\mu\nu}\left(-1+\frac{\eta^2k^2}{(k \cdot \eta)^2}\right)+\frac{k^{\mu}\eta^{\nu}+k^{\nu}\eta^{\mu}}{(k \cdot \eta)}-\frac{k^{\mu}k^{\nu}\eta^2+\eta^{\mu}\eta^{\nu}k^2}{(k \cdot \eta)^2}~,
	\end{eqnarray}
	if the photon is on the mass shell, $k^2=0$, and the Ward identity is used, $k^{\mu}$ and $k^{\nu}$ terms vanish in the amplitude of monopoles on the mass shell, and hence $\sum_{\gamma's~pol.}\epsilon_{m}(k)^{*\mu}\epsilon_m (k)^{\nu}=-g^{\mu\nu}$.
	Therefore, the expression of $\sigma(m+\overline{m} \to \gamma\gamma)$ is obtained from $\sigma(e^{+}+e^{-} \to \gamma \gamma)$, by replacing the mass $m_e$ by $m$ and the coupling constant $e$ by $g$.
	
	Now we can obtain the full form of the two photon decay rate as,
	\begin{eqnarray}
	\Gamma(M \to 2\gamma)=\Gamma_{2\gamma}=16\pi  \frac{\alpha_g^2 }{M^2}\frac{\int_{0}^{R}\vert \psi(r)\vert^2 dr}{\frac{4}{3}\pi R^3}~,
	\label{decaywidthgg}
	\end{eqnarray}
	which is the formula valid at low energy, in the case of $\sqrt{\hat{s}}=M  \approx 2m $. This approximation will be used in our subsequent analysis. 
	
	Due to the large coupling associated with the monopole coupling, this decay width can be very large depending upon the relation between the average probability density and the monopolium mass. We will consider as a general requirement throughout our work that the total decay width is less than $ 10\% $ of the corresponding monopolium mass, in order to not strongly violate the narrow width approximation. 
	
\begin{figure}
		\centering
		\includegraphics[width=0.5\textwidth]{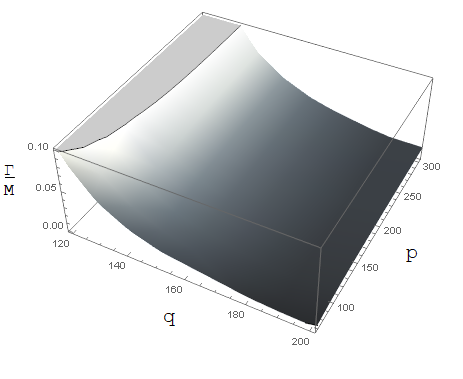}
		\caption{Ratio of the diphoton decay width and monopolium mass, for varying $ p $ and $ q $.}
		\label{2gdecaywidth}
	\end{figure}

	In Figure \ref{2gdecaywidth} we depict the dependence of the two photon decay width to monopolium mass ratio for a range of $ p $ and $ q $ parameters. The decrease of this ratio with increasing $ q $ is attributable to the enlargement of the monopolium mass and decrease in the average probability density. Values of $ q\lesssim 130$ lead to large decay widths for $ p>40 $, so we will consider $ q>130 $ for the analysis of the lowest energy level monopolum states.

	We must also consider the  contribution to the decay rate coming from the emission of multi-photons due to the monopoles strong coupling to photons. The total decay rate will become,
	\begin{eqnarray}
	\Gamma_\mathrm{M}&=&\sum_{n=\mathrm{even}}\Gamma(M \to n \gamma)~,
	\end{eqnarray}
	to calculate this total decay width we will follow \cite{Fanchiotti:2017nkk}, which utilises an  analogy to multiphoton positronium decays to estimate the branching ratios of higher multiplicity photon final states. In the next section we will confine ourselves to the two photon final state only, with the subsequent section introducing higher multiplicity photon decays and their collider phenomenological implications.


	\section{Production Cross Section and LHC Constraints on Monopoles from Monoplium Diphoton Decays}
	\label{twophoton_only}

	Now that we have obtained the states that may contribute to resonance like collider signatures, we can derive the current constraints on the monopolium model parameters and corresponding monopole mass. The current strongest cross section limits in the diphoton channel are provided by the ATLAS collaboration with $ \sim  $139 fb$ ^{-1} $ of data \cite{Aad:2021yhv}, and CMS collaboration with $ \sim  $39 fb$ ^{-1} $ at 13 TeV \cite{Sirunyan:2018wnk}. The derived monopole mass constraints can then be compared to those derived from the MoEDAL experiment which has been designed to search for pair produced monopoles.
	
	We will not consider states for which $ \Gamma_M>0.1 M $, as the smeared resonance structures exhibited by these states are weakly constrained by current experimental bounds. Despite this, for cases where the energy difference between the energy levels of the monopolium and the total decay widths are of comparable size, it may be important to consider possible interference effects. To calculate these effects would require analysis of the many possible states, which is beyond the scope of the current analysis. Our approximate assumption that these interference effects may not be detrimental will be discussed below.

	In calculating the predicted diphoton cross sections, we will begin by following the works of \cite{Epele:2008un}\cite{Epele2},
	\begin{equation}
	\hat{\sigma}_{\gamma \gamma} (\hat{s})=\frac{4 \pi}{\hat{s}} \frac{M^{2} \Gamma_{2\gamma} \Gamma_{M}}{\left(\hat{s}-M^{2}\right)^{2}+M^{2} \Gamma_{M}^{2}}~,
	\end{equation}
	where the total decay width includes the sum of all possible contributions, including multiphoton channels which will be discussed in the next section.
	
	Photon fusion processes are made up of three key processes, inelastic, semi-elastic and elastic, which must each be included to obtain the total predicted cross section. From \cite{Drees:1994zx} we can compute the inelastic contribution to the production cross-section, that in which neither of the protons remains intact after exchange of a photon. This is given by the following integral,
	\begin{equation}
	\begin{aligned} \sigma_{p p}^{\text{inel.}}(s)=& \int_{s_{th} / s}^{1} d x_{1} \int_{s_{th} / s x_{1}}^{1} d x_{2} \int_{s_{th} / s x_{1} x_{2}}^{1} d z_{1} \int_{s_{th} / s x_{2} x_{2} z_{1}}^{1} d z_{2} \frac{1}{x_{1}} F_{2}^{p}\left(x_{1}, Q^{2}\right) \\ & \cdot \frac{1}{x_{2}} F_{2}^{p}\left(x_{2}, Q^{2}\right) f_{\gamma}\left(z_{1}\right) f_{\gamma}\left(z_{2}\right) \hat{\sigma}_{\gamma \gamma}\left(x_{1} x_{2} z_{1} z_{2} s\right)~, \end{aligned}
	\end{equation}
	where the energy of the subprocess is given by $\hat{s}=x_1 x_2 z_1 z_2 s$, $\sqrt{s_{th}}=M$ is the threshold centre of mass energy, and the factors $F_2^p(x,Q^2)$ represent the deep inelastic proton structure function. The value of $Q^2$ is chosen throughout to be $\hat{s}/4$. Subsequently, we use the NNPDF2.3QED package to compute the partonic densities in the proton, in order to calculate the cross section. 
	
	In the semi-elastic process, only one of the protons remains intact after photon exchange. The associated production cross section is given by,
	\begin{equation}
	\begin{aligned} \sigma_{p p}^{\text {semi-el.}}(s)=& 2 \int_{s_{th} / s}^{1} d x_{1} \int_{s_{th} / s x_{1}}^{1} d z_{1} \int_{s_{th} / s x_{1} z_{1}}^{1} d z_{2} \frac{1}{x_{1}} F_{2}^{p}\left(x_{1}, Q^{2}\right) \\ & \cdot f_{\gamma}\left(z_{1}\right) f_{\gamma / p}^{e l .}\left(z_{2}\right) \hat{\sigma}_{\gamma \gamma}\left(x_{1} z_{1} z_{2} s\right) ~,\end{aligned}
	\end{equation}
	where $\hat{s}=x_{1} z_{1} z_{2} s$. For the photon spectrum $f_{\gamma}(z)$, we use
	\begin{equation}
	f_{\gamma}(z) =\frac{\alpha}{2 \pi} \frac{\left(1+(1-z)^{2}\right)}{z} \ln \left(\frac{Q_{1}^{2}}{Q_{2}^{2}}\right)~,
	\end{equation}
	where $Q_1^2=\hat{s}/4-M^2/4$ and $Q_2^2=1$ GeV$^2$.

	To calculate the elastic photon spectrum we use an approximate analytic form utilised in \cite{Drees:1994zx}, which are based upon the Weizs\"{a}cker-Williams approximation \cite{vonWeizsacker:1934nji,Drees:1988pp,Kniehl:1990iv}. It is of the following form,
	\begin{equation}
	f_{\gamma / p}^{e l .}(z)=\frac{\alpha}{2 \pi z}\left(1+(1-z)^{2}\right)\left[\ln A-\frac{11}{6}+\frac{3}{A}-\frac{3}{2 A^{2}}+\frac{1}{3 A^{3}}\right]~,
	\end{equation}
	where
	\begin{equation}
	A=1+\frac{0.71(\mathrm{GeV})^{2}}{Q_{\text {min}}^{2}}~,
	\end{equation}
	and
	\begin{equation}
	\begin{aligned} Q_{\min }^{2} &=-2 m_{p}^{2}+\frac{1}{2 s}\left[\left(s+m_{p}^{2}\right)\left(s-z s+m_{p}^{2}\right)\right.\\ &-\left(s-m_{p}^{2}\right) \sqrt{\left(s-z s-m_{p}^{2}\right)^{2}-4 m_{p}^{2} z s} ]~, \end{aligned}
	\end{equation}
	where in the limit $s \gg m^2_p$, we can approximate this as
	\begin{equation}
	Q_{\min}^2 \simeq \frac{m^2_p z^2}{1-z}~.
	\end{equation}
	Finally, the elastic contribution, where both protons remain intact after exchanging a photon, is given by
	\begin{equation}
	\sigma_{p p}^{\text{el.}}(s)=\int_{s_{th} / s}^{1} d z_{1} \int_{s_{th} / z_{1} s}^{1} d z_{2} \quad f_{\gamma / p}^{e l .}\left(z_{1}\right) f_{\gamma / p}^{e l .}\left(z_{2}\right) \hat{\sigma}_{\gamma \gamma}\left(z_{1} z_{2} s\right)~,
	\end{equation}
	where $\hat{s}=z_{1} z_{2} s$.
	
	The total monopolium production cross section will be given by the sum of these three contributions. From this we can now begin our analysis of the diphoton channel predictions for our model at the LHC. To start we will consider the scenario in which the monopolium only has one decay path, into two photon, and consider only the lowest energy eigenvalue for given $ p $ and $ q $ values. Thus we use the two photon decay width given in Eq. (\ref{decaywidthgg}) as the total decay width, and consider only $ p $ and $ q $ values within the range $ p\in[60,2000] $ and $ q\in[130,200] $.

	\subsection{Predicted Cross Section to Diphoton Final State }
	We can now calculate the predicted diphoton cross sections for our monopolium model at the 13 TeV LHC searches. In calculating these we have included the elastic, inelastic and semi elastic contributions to the photon fusion processes. These results can then be compared to the latest constraints determined by the ATLAS and CMS collaborations, to see which parameter regions of our model have already been probed. 
	The diphoton channel  constraints we utilise are the $ 95\% $ CL constraints  from ATLAS and CMS, across the $ S=0 $ state mass ranges $ [500,3000] $ GeV and $ [2250,4600] $ GeV, respectively. The range of masses considered by the ATLAS Collaboration, $ [500,3000] $ GeV, is smaller than that given by CMS, $ [500,4600] $ GeV. Although ATLAS utilises $  $139 fb$ ^{-1} $ \cite{Aad:2021yhv}, compared to $  $39 fb$ ^{-1} $ at 13 TeV \cite{Sirunyan:2018wnk} for CMS, hence providing more stringent constraints for the mass range  $ [500,3000] $ GeV. Thus, we include both analyses in order to extend the monopolium masses probed to a maximum of 4.6 TeV.  Given the prevalence of large decay widths in our formulation, it is important to note the constraints applied by the ATLAS collaboration can probe widths as large as $ 10\% $ of the monopolium mass, see Table \ref{Ch3repres}, while the results we quote form the CMS collaboration are for $ \Gamma_M/M\sim 5.6 \% $.

	\begin{table}[ht]
		\label{Ch3smt}
		\centering
		\begin{tabular}{|c|c c |}
			\hline
			M & 400 GeV & 2800 GeV\\\hline
			NWA & 1.1 fb & 0.03 fb\\
			$ \Gamma_M/M=  2\% $& 2.5 fb & 0.03 fb\\
			$ \Gamma_M/M=  6\%$ & 4.4 fb & 0.03 fb\\
			$ \Gamma_M/M= 10\%$ & 8.3 fb & 0.03 fb\\ \hline
		\end{tabular}
		\caption{Current ATLAS diphoton cross section constraints for both the Narrow Width Approximation (NWA), and broad width resonances \cite{Aad:2021yhv}.}
		\label{Ch3repres}
	\end{table}
	
\begin{figure}[h]
		\centering
		\begin{subfigure}
			\centering
			\includegraphics[width=0.42\textwidth]{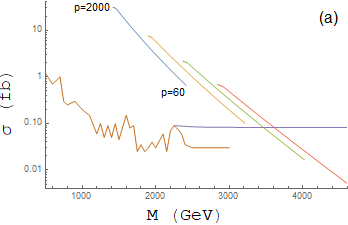}
		\end{subfigure}
		\hfill
		\begin{subfigure}
			\centering
			\includegraphics[width=0.42\textwidth]{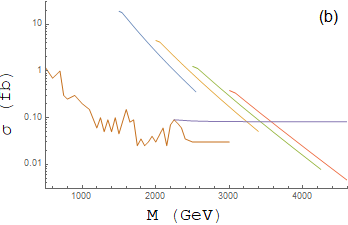}
		\end{subfigure}
		\\
		\begin{subfigure}
			\centering
			\includegraphics[width=0.42\textwidth]{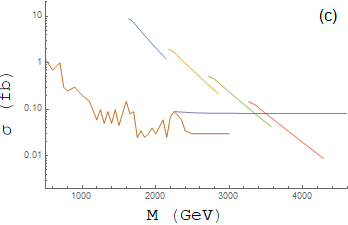}
		\end{subfigure}
		\hfill
		\begin{subfigure}
			\centering
			\includegraphics[width=0.42\textwidth]{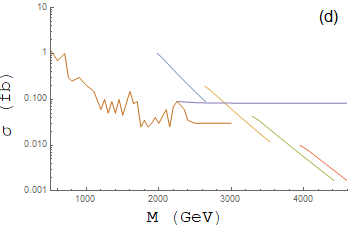}
		\end{subfigure}
		\caption{	The predicted diphoton cross-sections at 13 TeV LHC for the monopolium ground state ($ n=1 $), which is assumed to solely decay to two photons $ \Gamma_M = \Gamma_{2\gamma} $ with $ \Gamma_M <0.1 M $. The monopole masses considered are $ m= $1500 (blue), 2000 (yellow), 2500 (green), and 3000 GeV (red), for fixed $ q = $ (a) $ 130$, (b) 137, (c) 150, and (d) 200. The results across a range of $ p $ are used, $ p\in[60,2000] $, with (a) labelled with corresponding $ p $ values at the extrema,  which are consistent across each figure. The latest $ 95\% $ CL constraints from the CMS (purple) \cite{Sirunyan:2018wnk} and ATLAS (brown) \cite{Aad:2021yhv} collaborations  are included.}
		\label{cs2}
	\end{figure}

	The dependence of the predicted diphoton cross section on $ p $ and $ q $ is depicted in Figure \ref{cs2}. In each figure we fix $ q $ to different values and vary $ p $ within $ p\in[60,2000] $. The variation of $ p $ seen here indicates a continuous decrease in the predicted cross section with decreasing $ p $, and equivalently increasing monopolium mass. Larger $ p $ values decrease the monopolium mass until the $ p $ dependent term in the monopolium mass becomes negligible, after which further decreases in $ p $ have minimal effects. On the other hand, increasing $ q $ values lead to larger monopolium masses for given $ p $ and $ m $.

\begin{figure}
				\centering
		\begin{subfigure}
			\centering
			\includegraphics[width=0.45\textwidth]{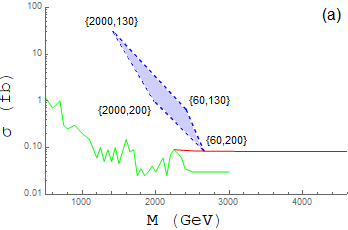}
		\end{subfigure}
		\hfill
		\begin{subfigure}
			\centering
			\includegraphics[width=0.45\textwidth]{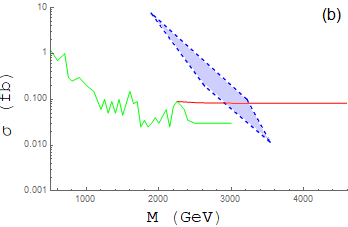}
		\end{subfigure}
		\begin{subfigure}
			\centering
			\includegraphics[width=0.45\textwidth]{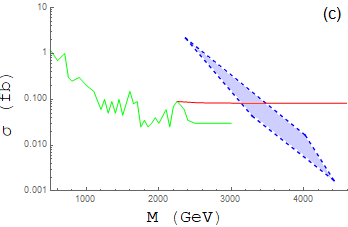}
		\end{subfigure}
		\hfill
		\begin{subfigure}
			\centering
			\includegraphics[width=0.45\textwidth]{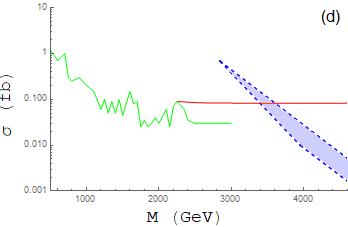}
		\end{subfigure}
		\begin{subfigure}
			\centering
			\includegraphics[width=0.45\textwidth]{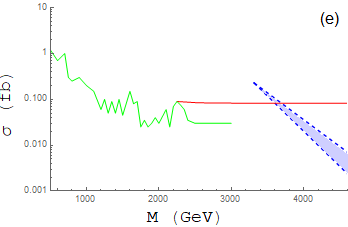}
		\end{subfigure}
		\hfill
		\begin{subfigure}
			\centering
			\includegraphics[width=0.45\textwidth]{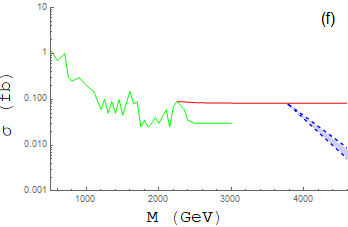}
		\end{subfigure}
		\caption{The predicted diphoton cross-sections at 13 TeV LHC for the monopolium ground state ($ n=1 $), which is assumed to solely decay to two photons $ \Gamma_M = \Gamma_{2\gamma} $ with $ \Gamma_M <0.1 M $. The monopole masses considered are $m=~\textrm{(a)}~1500,~\textrm{(b)}~2000,~\textrm{(c)}~2500,~\textrm{(d)}~3000,~\textrm{(e)}~3500,~\textrm{(f)}~4000$ GeV respectively, and the latest $ 95\% $ CL LHC constraints from the  CMS (red) \cite{Sirunyan:2018wnk} and ATLAS (green) \cite{Aad:2021yhv} collaborations are included. The ATLAS constraints have utilised the narrow width approximation, while the CMS constraints assume $ \Gamma_M/M\sim 5.6 \% $. The blue regions subtend the range of predicted cross sections for parameters $ p\in [60,2000] $ and $ q\in [70,200] $, with (a) including $ \{p,q\} $ labelling of extrema, which are consistent across each figure. }
		\label{2photon}
	\end{figure}

	By comparing different fixed $ q $ values we can see the monopole mass constraints across the range of possible $ p $. The variation observed for the full range of parameters makes it difficult to apply explicit limits on the allowed monopole masses. In Figure \ref{2photon}, we show the allowed parameter regions for fixing only the monopole mass. The parameters $ p $ and $ q $ are varied across a range of values for which $ \Gamma_{2\gamma}< 0.1 M $, namely  $ p\in [60,2000] $ and $ q\in [130,200] $. Due to the continuous behaviour observed for varying $ p $ and $ m $ for a fixed $ q $, we obtain a polygon which subtends the region of predicted cross sections for a given monopole mass, within our model. 
	
	In the context of our monopolium construction, we can thus constrain monopole masses to be greater than 1500 GeV for the entire model parameter space. Constraints can be applied on different sets of input parameters up to 4000 GeV. The most theoretically motivated choice of the $ q $ parameter, is likely $ q\sim 137 $ due to the relation to the classical monopole radius. For this value of $ q $ the monopole mass can be constrained up to $ \sim 3500 $ GeV, which is larger than the current constraints from the MoEDAL experiment.

	The shape of the predicted cross section region is approximately conserved with changing monopole mass, which is expected from the factoring out of $ m $ dependence in the energy eigenvalue calculation. The region is then defined by the range of $ p $ and $ q $ values that are taken. Once $ p \gg q$, there is no increase in the size of this region because the $ p $ contribution to the monopolium binding energy gets increasingly suppressed. An increase in the region is only possible through considering smaller $ p $ and $ q $ values, but these have minimum values fixed by the requirement that the monopolium mass satisfies $ 0<M<2m $. In the next section, we include model parameters with smaller $ q $ values that still maintain $ M>0 $, for which the lowest energy eigenvalues that exhibited large decay widths. We do this by considering higher energy levels of the bound states, which can have decay widths satisfying $ \Gamma<0.1 M $.

	\subsection{Including Higher Energy Eigenstates }
	
	For the model parameters within the range $ 70<q<130 $ the lowest energy level consists of a total decay width that is too large to be probed by current collider results, which assume narrower resonances. In order to test this region of parameter space we consider the lowest energy state for which $ \Gamma_M=\Gamma_{2\gamma}<0.1 M $ is satisfied. From the production cross section extracted from these energy eigenstates we can then attempt to constrain them.

\begin{figure}[h!]
		\centering
		\includegraphics[width=0.6\textwidth]{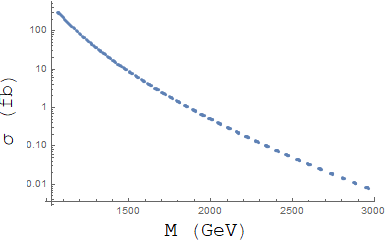}
		\caption{Predicted two photon cross section for each energy level ($ n\geq 1 $) satisfying $ \Gamma_{M}/M<0.1 $, for the fixed parameters $ p=q=100 $ and $ m=1500 $ GeV. Assuming the monopolium decays solely to two photons $ \Gamma_M = \Gamma_{2\gamma} $.}
		\label{higher_ens}
	\end{figure}

	The justification for this choice can be seen in Figure \ref{higher_ens}, which depicts all the energy eigenvalues with $\Gamma_{M}<0.1 M $ for the $ q=p=100 $ model, and their predicted cross sections. It can be seen that the state with the largest cross section is that which has the lowest energy eigenvalue, and hence smallest monopolium mass, with all subsequent states are suppressed. This suppression is caused by the combination of larger monopolium masses, and the decrease of the probability density factor $ \rho $ for increasingly excited energy states.

	 Although it may be possible to probe other states in this series, we will consider only the one with the largest cross section. One possible concern with this analysis is the potential for interference effects between the different energy eigenstates. Although this could be an important effect for the large decay widths considered here, for simplicity we assume the relative cross sections is sufficiently suppressed for the higher energy states that there contributions can be ignored. A full analysis of the possible implications of this interference is beyond the scope of the current work, with further investigation of interest in future work.

	\begin{figure}
		\centering
		\begin{subfigure}
			\centering
			\includegraphics[width=0.38\textwidth]{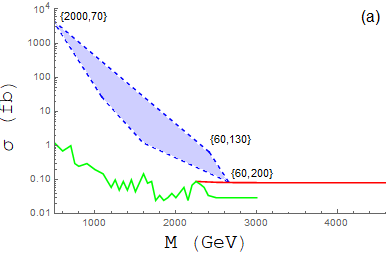}
		\end{subfigure}
		\hfill
		\begin{subfigure}
			\centering
			\includegraphics[width=0.38\textwidth]{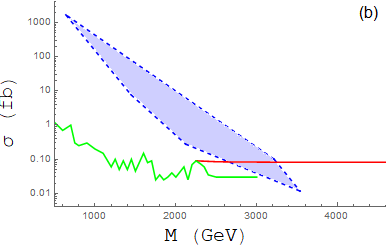}
		\end{subfigure}
		\begin{subfigure}
			\centering
			\includegraphics[width=0.38\textwidth]{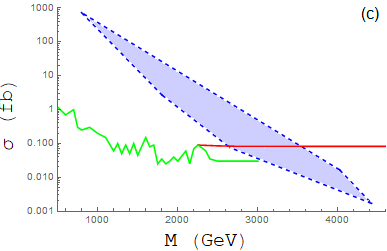}
		\end{subfigure}
		\hfill
		\begin{subfigure}
			\centering
			\includegraphics[width=0.38\textwidth]{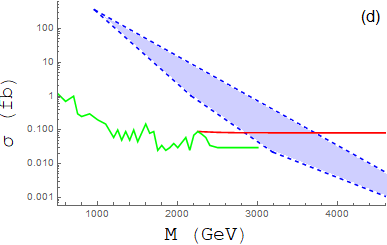}
		\end{subfigure}
		\begin{subfigure}
			\centering
			\includegraphics[width=0.38\textwidth]{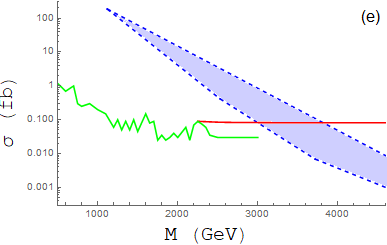}
		\end{subfigure}
		\hfill
		\begin{subfigure}
			\centering
			\includegraphics[width=0.38\textwidth]{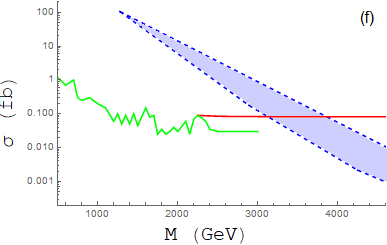}
		\end{subfigure}
		\begin{subfigure}
			\centering
			\includegraphics[width=0.38\textwidth]{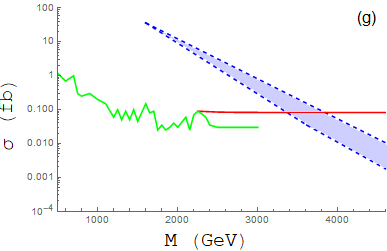}
		\end{subfigure}
		\hfill
		\begin{subfigure}
			\centering
			\includegraphics[width=0.38\textwidth]{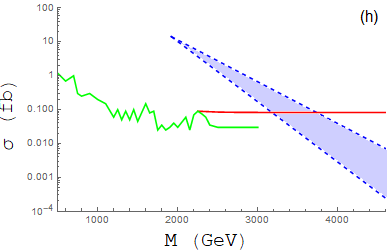}
		\end{subfigure}
		\caption{	The predicted diphoton cross-sections at 13 TeV LHC for the monopolium ground state ($ n=1 $) and excited states ($ n = 2 $, 3,...), assuming decay solely into two photons $ \Gamma_M = \Gamma_{2\gamma} $ with $ \Gamma_M<0.1 M $. The monopole masses considered are  $ m=~\textrm{(a)}~1500 ,~\textrm{(b)}~2000,~\textrm{(c)}~2500,~\textrm{(d)}~3000,~\textrm{(e)}~3500,~\textrm{(f)}~4000,~\textrm{(g)}~5000,~\textrm{(h)}~6000 $ GeV respectively, and the latest $ 95\% $ CL LHC constraints from the  CMS (red) \cite{Sirunyan:2018wnk} and ATLAS (green) \cite{Aad:2021yhv} collaborations are included. The blue regions subtend the range of predicted cross sections for parameters $ p\in [60,2000] $ and $ q\in [130,200] $, with (a) including $ \{p,q\} $ labelling of extrema, which are consistent across each figure.}
		\label{2photonneq}
	\end{figure}

	\begin{figure}
		\centering
		\begin{subfigure}
			\centering
			\includegraphics[width=0.38\textwidth]{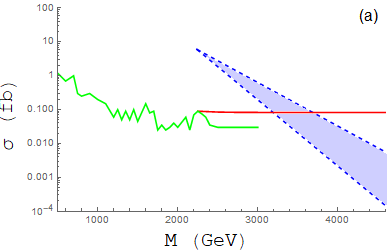}
		\end{subfigure}
		\hfill
		\begin{subfigure}
			\centering
			\includegraphics[width=0.38\textwidth]{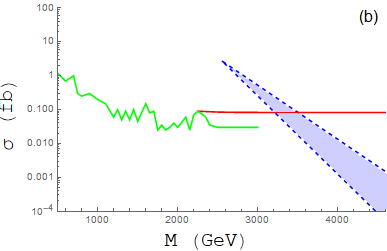}
		\end{subfigure}
		\begin{subfigure}
			\centering
			\includegraphics[width=0.38\textwidth]{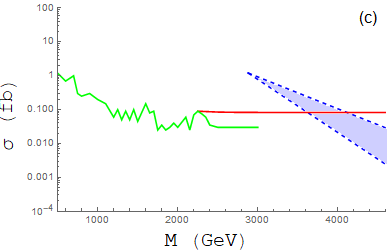}
		\end{subfigure}
		\hfill
		\begin{subfigure}
			\centering
			\includegraphics[width=0.38\textwidth]{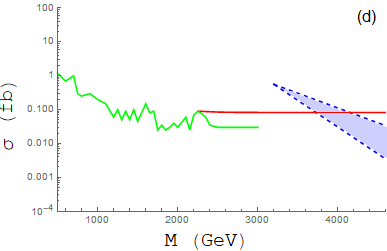}
		\end{subfigure}
		\begin{subfigure}
			\centering
			\includegraphics[width=0.38\textwidth]{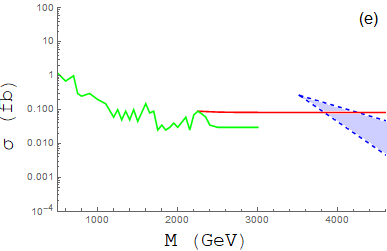}
		\end{subfigure}
		\hfill
		\begin{subfigure}
			\centering
			\includegraphics[width=0.38\textwidth]{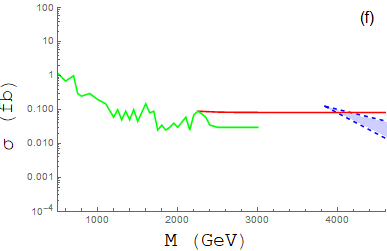}
		\end{subfigure}
		\begin{subfigure}
			\centering
			\includegraphics[width=0.38\textwidth]{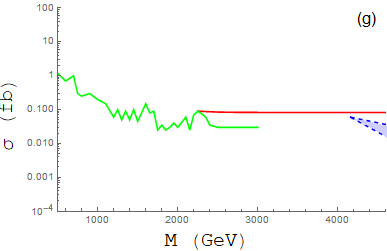}
		\end{subfigure}
		\hfill
		\begin{subfigure}
			\centering
			\includegraphics[width=0.38\textwidth]{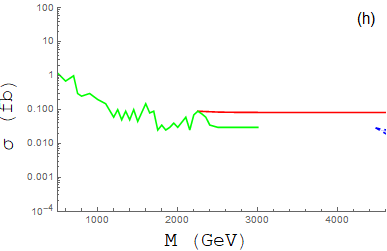}
		\end{subfigure}
		\caption{The predicted diphoton cross-sections at 13 TeV LHC for the monopolium ground state ($ n=1 $) and excited states ($ n = 2 $, 3,...), assuming decay solely into two photons $ \Gamma_M = \Gamma_{2\gamma} $ with $ \Gamma_M<0.1 M $. The monopole masses considered are  $ m=~\textrm{(a)}~7000 ,~\textrm{(b)}~8000,~\textrm{(c)}~9000,~\textrm{(d)}~10000,~\textrm{(e)}~11000,~\textrm{(f)}~12000,~\textrm{(g)}~13000,~\textrm{(h)}~14000 $ GeV respectively, and the latest $ 95\% $ CL LHC constraints from the  CMS (red) \cite{Sirunyan:2018wnk} and ATLAS (green) \cite{Aad:2021yhv} collaborations are included. The blue regions subtend the range of predicted cross sections for parameters $ p\in [60,2000] $ and $ q\in [70,200] $.}
		\label{2photonneq3}
	\end{figure}

	By opening our analysis to this higher energy states we can probe smaller values of $ q $, namely we consider $ 70<q<130 $, in addition to $ n=1 $ states for $130<q<200 $. As can be seen in Eq. (\ref{mon_mass}), such model parameter selections leads to larger negative binding energies, and thus smaller monopolium masses for a given monopole mass. This gives an opportunity to probe even larger monopole masses than would otherwise be possible be monopole-anti-monopole pair production.

	In Figures \ref{2photonneq} and \ref{2photonneq3}, we provide the expected cross sections for the monopolium diphoton decays for the model parameter range $ q\in[70,200]  $ and $ p\in[60,2000] $. The monopole mass is chosen for various values between 1500 GeV and 14000 GeV for illustrative purposes.  By comparing these results to those in Figure \ref{2photon}, for the ground state case, we see that including excited states ($ n=2 $, 3,...) has resulted in enlarging the predicted cross section region. This is due to the inclusion of monopolium models consisting of $ q<130 $ parameter values, which were originally excluded as their ground state decay widths violated our requirement that $ \Gamma_M<0.1M $. The excited energy states of $ q<130 $ monopoliums exhibit larger cross sections and smaller bound state masses, than the ground states of $ q>130 $ models.  Furthermore, where in the $ n=1 $ case, no parameter constraints could be found for monopole masses greater than 4000 GeV, in the $ n\geq 1 $ case, it is possible to constrain monopole masses as large as 12000 GeV  from Figure  \ref{2photonneq3}, for certain input parameter. This points to the key advantage and motivation of considering the formation of monopole-anti-monopole bound states.

	
	\section{Multi-photon Annihilation of Monopolium and Branching Ratio to Diphoton}
	\label{multiphoton}
	
	In our analysis so far, we have neglected the possibility of other decay channels existing beyond the expected diphoton decays. Issues with monopoles arise when calculating the total decay width due to the non-perturbative nature of their couplings, with higher order photon decays can become enhanced leading to very large total widths. In our previous work, we attempted to control this divergent behaviour but the resultant total decay widths were well beyond the mass of the monopolium itself, making it difficult to derive any meaningful phenomenological constraints. Here we will consider the possibility of making an analogy between the monopolium multiphoton decays and those of the positronium bound states, which was recently suggested in Ref. \cite{Fanchiotti:2017nkk}. By making this connection the divergence of the full width is constrained by phase space suppression effects. The model tends to lead to the prediction of significant multiphoton final states with multiplicity greater than two, as suggested in our previous work. This can lead to interesting phenomenology, as even if the production rate of monopolium states is large their decays may be hidden from the diphoton channel, with the majority of decays occurring into higher multiplicity final states, which are difficult to constrain at present colliders. One may be able to obtain tentative bounds on these from two photon plus missing energy searches, but a thorough analysis is needed. This is  strongly motivated by the expected large signals that can be produced by monopolium decays. The branching ratio model suggested in Ref. \cite{Fanchiotti:2017nkk} may be valued more on a qualitative than quantitative basis, but we will consider the implications for our monopolium construction here, under the assumption that this positronium analogy is approximately valid.

	\subsection{Branching Ratio to Higher Multiplicity Photon Decays}
	
	In \cite{Fanchiotti:2017nkk}, the ratios of the higher multiplicity photon final states were matched to those of the positronium states, to obtain a numerical approximation for higher $ n $ multiplicities. The resultant ratio of the rate of multiphoton emission is,
	\begin{equation}
	\frac{\Gamma_{2 n\gamma}}{\Gamma_{2\gamma}}=\left(\frac{1}{2}\right)^{2 n-2}\left(\frac{\alpha_g}{\pi}\right)^{2 n-2}\left(\frac{M}{2 m}\right)^{4 n-4} \frac{2 n !}{2 !(2 n-1) !(2 n-2) !}~,
	\label{ratio}
	\end{equation}
	where $\Gamma_{2n\gamma}$ is the decay width of monopolium to $n$ photon pairs, and $\alpha_g = 1/\alpha =137$ for monopoles of charge $ g $. Note that, this approximate relation assumes that the monopole loop can be shrunk to a point due to the monopole mass being very heavy, and hence that the monopolium and multi-photons interact via a contact interaction. 
	
	The total decay width is then given by the sum over all final state pairs of photons,
	\begin{equation}
	\Gamma_M = \sum_n \Gamma_{2n\gamma}~,
	\end{equation}
	which allows us to write,
	\begin{equation}
	\frac{\Gamma_M}{\Gamma_{2\gamma}}=\sum_n \frac{n }{(2 n-2)! }\left(\frac{137 M^2}{8\pi m^2}\right)^{2 n-2}~,
	\end{equation}
	from which we can derive at what monopolium masses each higher multiplicity final state can be expected to dominate the total decay width,
	\begin{equation}
	\small
	\frac{\Gamma_M}{\Gamma_{2\gamma}} \approx \underbrace{1}_{n_\gamma=2}+\underbrace{\left(\frac{M}{0.43 m}\right)^{4 }}_{n_\gamma=4}+\underbrace{\left(\frac{ M}{0.56 m}\right)^{8}}_{n_\gamma=6}+\underbrace{\left(\frac{ M}{ 0.66 m}\right)^{12}}_{n_\gamma=8} + \underbrace{\left(\frac{ M}{ 0.75 m}\right)^{16}}_{n_\gamma=10}+...+\underbrace{\left(\frac{ M}{ 1.04 m}\right)^{32}}_{n_\gamma=18}+...
	\label{ratio2}
	\end{equation}
	
	From Eqs. (\ref{ratio}) and (\ref{ratio2}), we see that the decay to higher multiplicities $ n $ is favoured for larger monopolium masses, relative to the monopole mass.  This has implications of varying importance for the diphoton branching ratio dependent upon the $ p $ and $ q $ values considered.  Interestingly, this behaviour leads to an enhancement of the total decay width for increasingly higher order energy states, which is counter to the decrease observed in solely the two photon decay width case discussed in Section \ref{twophoton_only}. This reduces the number of eigenstates that satisfy $ \Gamma_\textrm{M}<0.1M $, while simultaneously suppressing the branching ratios of higher energy states to diphotons. The predicted diphoton cross section for the lowest energy state satisfying $ \Gamma_\textrm{M}<0.1M $ will thus dominate over the higher energy states. 
	
	\begin{figure}[h!]
		\centering
		\begin{subfigure}
			\centering
			\includegraphics[width=0.45\textwidth]{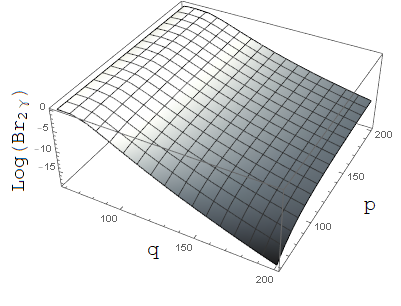}
		\end{subfigure}
		\hfill
		\begin{subfigure}
			\centering
			\includegraphics[width=0.45\textwidth]{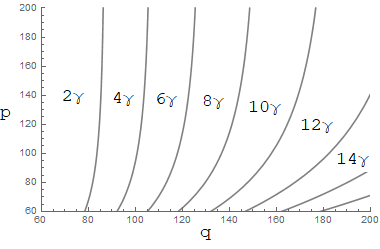}
		\end{subfigure}
		\caption{Branching ratio to two photons (left) as a function of $ q $ and $ p $, for the corresponding lower bound on the monopolium mass $ M_0 $. The right figure depicts the dominant photon multiplicity final state for given $ p $ and $ q $ values. }
		\label{Brs}
	\end{figure}

	In Figure \ref{Brs}, we can see how changing values of $ q $ and $ p $ effects the expected branching ratios to different multiplicity photon states. The key behaviour, is that for increasing values of $ q $ the decay width to two photons becomes rapidly suppressed. On the other hand, the dependence on $ p $ has a minor effect when $ p>q $, with minimal effect for larger $ p $ values. This can be deduced by considering the importance of the monopolium to monopole mass term in the decay width ratio equation, Eq. (\ref{ratio}), with the $ p $ and $ q $ dependence of the monopolium mass, see Eq. (\ref{mon_mass}). Constraints from collider searches on higher multiplicity photon final states have not yet been explored, so monopolium mass states with smaller $ |\Delta E| $ are hidden from current experimental probes. To illuminate their properties, specialised analysis of higher multiplicity final states needs to take place. There may also be many regions of the model parameter space for which meaningful constraints could be provided by multiple decay channels, with a discover in two channels being a possible smoking gun for a monopolium state. This provides strong motivation for dedicated searches in the mulitphoton final states.
	
	The inclusion of the extra possible decay channels means that the total decay widths are larger for all the monopolium states. In the previous section, we discussed the potential issues with constraining states exhibiting broad resonance structures. As such, in this scenario we will follow the methodology employed above in which we considered the higher energy eigenvalues, determining that state with the largest cross section while also maintaining $ \Gamma_\textrm{M}<0.1M $. Unlike the case with solely diphoton decay, the higher energy eigenvalues now have rapidly increasing decay widths due to the monopolium mass dependence in the branching ratio calculation. This means that there will be less states that satisfy the requirement of a small total decay width, particularly for larger $ q $ models. 
	
	\begin{figure}
		\centering
		\includegraphics[width=0.6\textwidth]{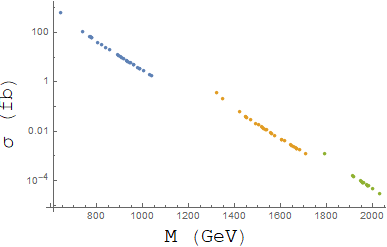}
		\caption{Predicted cross section for each excited monopolium state ($ n>1 $) satisfying $ \Gamma_{M}/M<0.1 $, when mulitphoton decays are included, i.e. $ \Gamma_M = \sum_n \Gamma_{2n\gamma} $. The parameters $ p=1000 $ and $ m=2000 $ GeV are fixed, with results for all allowed excited states given for $ q=80 $ (Blue), $ 100 $ (Orange), $ 120 $ (Green) respectively.}
		\label{highenmulti}
	\end{figure}
	
	In Figure \ref{highenmulti}, we show the dependence of the predicted cross section on the energy eigenvalues for three different sets of parameters, $ q=80 $, $ 100 $, $ 120 $, with $ p=1000 $ and $ m=2000 $ GeV fixed. This demonstrates the similar higher energy state cross section suppression to that depicted in Figure \ref{higher_ens}, but with fewer eigenvalues satisfying $ \Gamma_\textrm{tot}<0.1M $ due to the large decay width contribution to higher multiplicity decays. In what follows, we again assume that there is minimal interference effects between these states.

	\subsection{Constraints upon Inclusion of Multiphoton Decay Channels}
	
\begin{figure}
		\centering
		\begin{subfigure}
			\centering
			\includegraphics[width=0.38\textwidth]{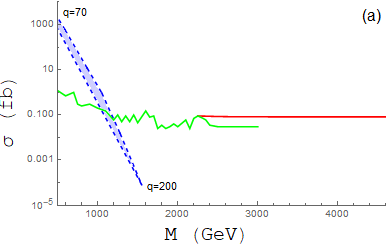}
		\end{subfigure}
		\hfill
		\begin{subfigure}
			\centering
			\includegraphics[width=0.38\textwidth]{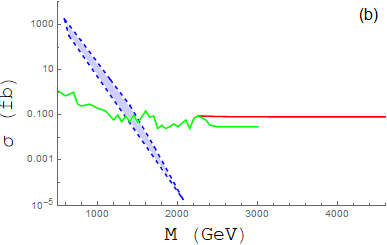}
		\end{subfigure}
		\begin{subfigure}
			\centering
			\includegraphics[width=0.38\textwidth]{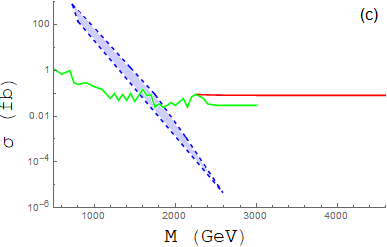}
		\end{subfigure}
		\hfill
		\begin{subfigure}
			\centering
			\includegraphics[width=0.38\textwidth]{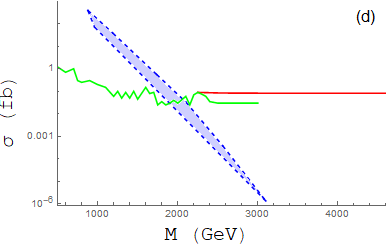}
		\end{subfigure}
		\begin{subfigure}
			\centering
			\includegraphics[width=0.38\textwidth]{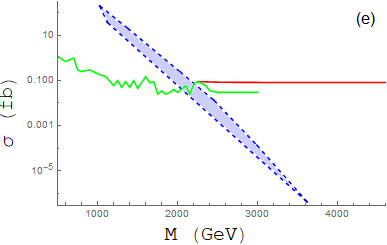}
		\end{subfigure}
		\hfill
		\begin{subfigure}
			\centering
			\includegraphics[width=0.38\textwidth]{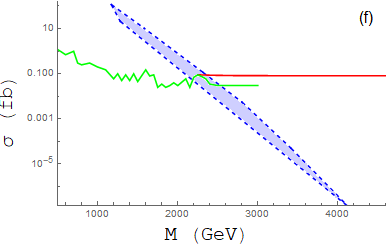}
		\end{subfigure}
		\begin{subfigure}
			\centering
			\includegraphics[width=0.38\textwidth]{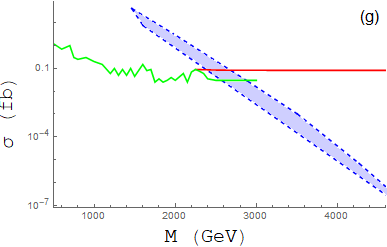}
		\end{subfigure}
		\hfill
		\begin{subfigure}
			\centering
			\includegraphics[width=0.38\textwidth]{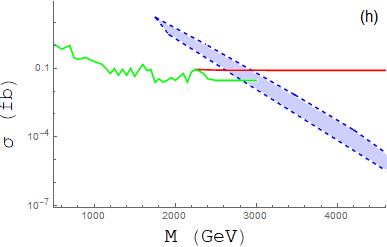}
		\end{subfigure}
		\caption{The predicted diphoton cross-sections at 13 TeV LHC for the monopolium ground state ($ n=1 $) and excited states ($ n = 2 $, 3,...), including multiphoton decay processes $ \Gamma_M = \sum_n \Gamma_{2n\gamma} $ with $ \Gamma_M<0.1 M $. The monopole masses considered are  $ m=~\textrm{(a)}~1500 ,~\textrm{(b)}~2000,~\textrm{(c)}~2500,~\textrm{(d)}~3000,~\textrm{(e)}~3500,~\textrm{(f)}~4000,~\textrm{(g)}~5000,~\textrm{(h)}~6000 $ GeV respectively, and the latest $ 95\% $ CL LHC constraints from the  CMS (red) \cite{Sirunyan:2018wnk} and ATLAS (green) \cite{Aad:2021yhv} collaborations are included. The blue regions subtend the range of predicted cross sections for parameters $ p\in [60,2000] $ and $ q\in [70,200] $, with (a) labelled with corresponding $ q $ values at the extrema, which are consistent across each figure.}
		\label{2photonneq1}
	\end{figure}

\begin{figure}
		\centering
		\begin{subfigure}
			\centering
			\includegraphics[width=0.38\textwidth]{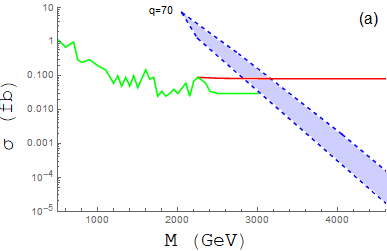}
		\end{subfigure}
		\hfill
		\begin{subfigure}
			\centering
			\includegraphics[width=0.38\textwidth]{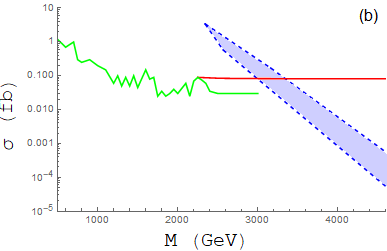}
		\end{subfigure}
		\begin{subfigure}
			\centering
			\includegraphics[width=0.38\textwidth]{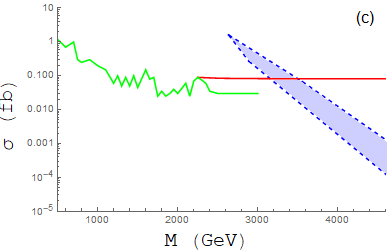}
		\end{subfigure}
		\hfill
		\begin{subfigure}
			\centering
			\includegraphics[width=0.38\textwidth]{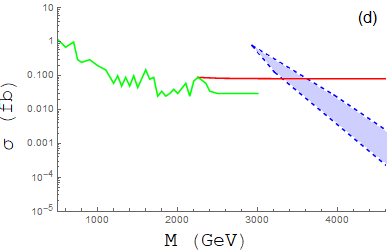}
		\end{subfigure}
		\begin{subfigure}
			\centering
			\includegraphics[width=0.38\textwidth]{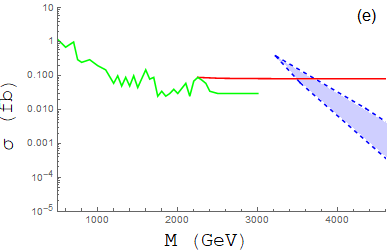}
		\end{subfigure}
		\hfill
		\begin{subfigure}
			\centering
			\includegraphics[width=0.38\textwidth]{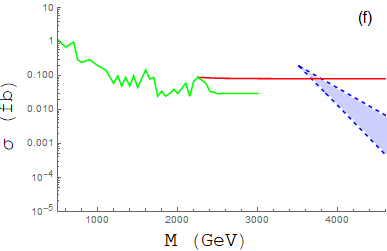}
		\end{subfigure}
		\begin{subfigure}
			\centering
			\includegraphics[width=0.38\textwidth]{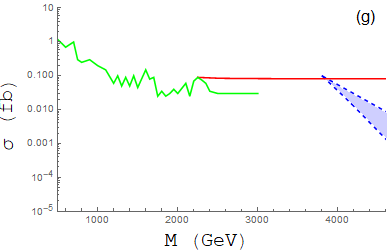}
		\end{subfigure}
		\hfill
		\begin{subfigure}
			\centering
			\includegraphics[width=0.38\textwidth]{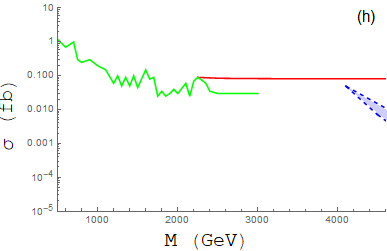}
		\end{subfigure}
		\caption{The predicted diphoton cross-sections at 13 TeV LHC for the monopolium ground state ($ n=1 $) and excited states ($ n = 2 $, 3,...), including multiphoton decay processes $ \Gamma_M = \sum_n \Gamma_{2n\gamma} $ with $ \Gamma_M<0.1 M $. The monopole masses considered are  $ m=~\textrm{(a)}~7000 ,~\textrm{(b)}~8000,~\textrm{(c)}~9000,~\textrm{(d)}~10000,~\textrm{(e)}~11000,~\textrm{(f)}~12000,~\textrm{(g)}~13000,~\textrm{(h)}~14000 $ GeV respectively, and the latest $ 95\% $ CL LHC constraints from the  CMS (red) \cite{Sirunyan:2018wnk} and ATLAS (green) \cite{Aad:2021yhv} collaborations are included. The blue regions subtend the range of predicted cross sections for parameters $ p\in [60,2000] $ and $ q\in [70,200] $, with (a) labelled with corresponding $ q $ values at the extrema, which are consistent across each figure.}
		\label{2photonneq2}
	\end{figure}

	In Figures \ref{2photonneq1} and \ref{2photonneq2}, we provide the expected cross sections for the monopolium diphoton decays for the model parameter range $ q\in[70,200]  $ and $ p\in[60,2000] $. The monopole mass is chosen for various values between 1500 GeV and 14000 GeV for illustrative purposes. The $ 95\% $ CL  diphoton channel  constraints from the LHC collaborations, ATLAS and CMS, are included across the $ S=0 $ state mass ranges $ [500,3000] $ GeV and $ [2250,4600] $ GeV, respectively. As in the previous scenarios, we can see that the monopole mass dependence has a small effect on the shape of the predicted cross section region, which instead is defined by the range of $ p $ and $ q $ values.

	Upon comparing these results to those depicted in Figures \ref{2photonneq} and \ref{2photonneq3}, it can be seen that the region describing each monopole masses  predicted cross section has been shifted, with the peak cross section remaining approximately the same, and the minimum being largely suppressed. Importantly, this leads to a narrowing of the predicted region, allowing easier separation of possible degeneracy between model parameters within uncertainty. Additionally, in allowing for higher multiplicity photon decays we can no longer fully rule out monopoles of mass 1500 GeV, as could be achieved when considering only diphoton decay. This is because those higher mass monopolium states now predominately decay into higher multiplicity states and as such are hidden from the diphoton channel. States that decay predominantly into higher multiplicity photon states require dedicated searches to be probed. This means their production by photon fusion is also suppressed. One may also consider multiphoton fusion processes to search for these particles, for example in heavy ion collisions. 
	
	Similar to the two photon only case, constraints can be applied on monopole mass models as high as 13 TeV, which is far beyond the current capabilities of dedicated monopole searches at the LHC. This is the key motivation of searching for monopoles through their bound state formation, and gives exciting possibilities for the monopole masses that could be probed by future colliders with larger centre of mass energies, such as 100 TeV.


\section{Conclusions}

The prospect of discovering the existence of monopoles could lead to many exciting phenomenological consequences, and also significant implications for cosmological observables, including Baryogenesis and gravitational waves. The strong coupling of monopoles leads to the possibility of monopole-anti-monopole pairs readily be produced in the form of bound states, monopolium. The large negative binding energy of monopolium states means that they can be copiously produced at energy scales well below that of the monopole mass itself. This can open the window to probing higher energy scales than could be reached by pair production alone. The monopolium states are expected to decay predominantly into multiphoton final states, which can provide a clean signal in comparison to that produced by monopole pair production. This possibility is a strong motivation for exploring further the physics of monopolium states.

In this work,  we have made enhancements to and expanded upon the phenomenological details of a monopolium construction we had previously proposed. This monopolium model utilised an analogy to the strong coupling regime description of quarkonium bound states, in which the bound state is formed non-perturbatively using a strong coupling expansion in lattice gauge theory. Improvements have been made in the calculation of the properties of the bound state, and the description of higher energy eigenvalues of the monopolium state. A new method has been used to control the divergence of total decay width upon including multiphoton decays, by utilising an analogy to multiphoton positronium decays that was  suggested by a recent work \cite{Fanchiotti:2017nkk}. These improvements have allowed a thorough analysis of the phenomenology of our monopolium construction at current colliders. For certain parameter regions of our model, we have derived constraints on the spin $ 1/2 $ monopole mass that exceed the strongest constraints available from the MoEDAL collaboration, namely $ m>2420 $ GeV \cite{Acharya:2019vtb}.

Here we have constrained our analysis to the para-monopolium ($ S=0 $) bound states with $ l=0 $, but the existence of ortho-monopolium states ($ S=1 $) is also expected. The signatures produced by decays to three photon final states would be an ideal complementary test to the para-monopolium decays, but at present there are minimal constraints in this decay channel \cite{Aaboud:2017lxm}. We have also assumed that the possible interference effects between energy eigenstates are minimal, which will require a more detailed analysis beyond the scope of the current work. Large areas of parameter space have been neglected due to the very large decay widths that they exhibit, $ \Gamma_{\textrm{tot}}>0.1 M $. Future theoretical and experimental techniques may make it possible to more easily probe these states. For example, see the recent progress in searches for $ t\bar{t} $ analyses which consider total decay widths as high as $ \Gamma_{\textrm{tot}}\sim 0.6 M $ \cite{Liu:2019bua}. Prior to the discovery of the Higgs boson, there was discussion of the idea of the stealthy Higgs \cite{Binoth:1999ay}. A large invisible Higgs width could lead to a very broad resonance allowing it to evade conventional methods. Such phenomenological implications considered therein could help guide how to consider such large multiphoton decay widths for the monopolium state. 

The building of new colliders such as the proposed 100 TeV collider \cite{Mangano:2017tke}, will be able to probe deeply into the parameter space of our model, possibly leading to the discovery of monopoles with masses of order $ \mathcal{O}(10) $ TeV. This in combination with new cosmological and gravitational wave experiments may provide new possibilities to search for the existence of monopoles.

\section*{Acknowledgements}
NDB is supported  by IBS under the project code, IBS-R018-D1, and by the World Premier International Research Center Initiative (WPI), MEXT, Japan. MT is supported by the grant ``AstroCeNT: Particle Astrophysics Science and Technology Centre" carried out within the International Research Agendas programme of the Foundation for Polish Science financed by the European Union under the European Regional Development Fund. KY is supported by the Chinese Academy of Sciences (CAS) President's International Fellowship Initiative under Grant No. 2020PM0018.

\end{document}